\documentstyle[12pt]{article}
\font\mygoth=eufm10 at 12pt
\newcommand\goth[1]{\hbox{\mygoth#1}}
\font\mybb=msbm10 at 12pt
\newcommand\bb[1]{\hbox{\mybb#1}}
\newcommand{\sect}[1]{\setcounter{equation}{0}\section{#1}}

\newcommand{\be}{\begin{equation}}
\newcommand{\en}{\end{equation}}
\newcommand{\bea}{\begin{eqnarray}}
\newcommand{\ena}{\end{eqnarray}}
\newcommand{\vs}[1]{\vspace{#1 mm}}
\newcommand{\nls}{non-linear Schr\"odinger }
\newcommand{\Wi}{\mbox{${\sf W}_{1+\infty}$}}
\newcommand{\ii}{r}
\newcommand{\sm}[2]{\frac{\mbox{\footnotesize #1}\vs{-2}}
                   {\vs{-2}\mbox{\footnotesize #2}}}
\newcommand{\PL}{Phys. Lett. }

\newcommand{\W}{\mbox{\sf W}}
\newcommand{\Q}{\mbox{$\tilde Q$}}
\newcommand{\h}{\mbox{$\tilde H$}}
\begin{document}
%
%
%
%
%
%
%
%
%
%
\renewcommand{\thefootnote}{\fnsymbol{footnote}}
\newpage
\setcounter{page}{0}
\pagestyle{empty}
\leftline{KCL-TH-93-1}
\leftline{January 1993}

\vs{20}

\begin{center}
{\Large {On the relation between integrability and
infinite-dimensional algebras\footnote{Talk given at the 1992
John Hopkins meeting, Goteborg, Sweden}}}\\[1cm]
{\Large M. D. Freeman and P. West}\\[0.5cm]
{\em  Department of Mathematics}\\
{\em King's College London, Strand, London WC2R 2LS}
\\[1cm]
\end{center}
\vs{15}

\centerline{ \bf{Abstract}}
We review our work on the relation between integrability
and infinite-dimensional
algebras.  We first consider the question of what
sets of commuting charges can be constructed from the
current of a \mbox{\sf U}(1) Kac-Moody algebra.
It emerges that there exists a set $S_n$ of such charges for each
positive integer $n>1$; the corresponding value of the central charge in the
Feigin-Fuchs realization of the stress tensor is $c=13-6n-6/n$.  The charges in
each series can be written in terms of the generators of an exceptional
\W-algebra.
We show that the \W-algebras that arise in this way are symmetries of
Liouville theory for special values of the coupling.
We then exhibit a relationship between the \nls equation and
the KP hierarchy.  From this it follows that there is a relationship between
the \nls equation and the algebra \Wi.  These examples provide evidence
for our conjecture that the phenomenon of integrability is intimately
linked with properties of infinite dimensional algebras.

\renewcommand{\thefootnote}{\arabic{footnote}}
\setcounter{footnote}{0}
\newpage
\pagestyle{plain}

%
%
%
\sect{Introduction}
A dynamical system with an infinite number of degrees
of freedom is said to be integrable if it possesses an
infinite number of conserved quantities in involution, i.e. for which the
Poisson bracket between any pair is zero.  Such systems have been
studied for many years and include the KdV equation,
\be
{\partial u \over \partial t} = {1\over 8} u^{\prime\prime\prime} -
{3\over 8} u u',
\en
the Liouville equation,
\be
\partial\bar\partial \phi - e^{-\phi} = 0,
\en
and the sinh-Gordon equation
\be
\partial\bar\partial \phi + 2 \sinh \phi = 0.
\en
These systems are in fact related to the Lie algebra $sl(2)$.
Corresponding to every Lie algebra there is an
integrable generalization of each of them \cite{DS,Wi},
namely the generalized KdV equations, Toda equations and affine Toda
equations respectively.

All of these equations can be written in Hamiltonian form, in which the
Hamiltonian generating the time evolution has the property that it commutes
with all of the conserved charges with respect to the Poisson bracket.
Each equation is therefore a member of an infinite hierarchy of equations,
obtained by using the conserved charges as Hamiltonians to define commuting
time evolutions for infinitely many times.

In another development, which has no obvious connection with
integrability, conformal field theories have been intensively studied
recently.  A theory that is Poincar\'e invariant possesses a conserved
energy-momentum tensor, and if such a theory is in addition conformally
invariant then the energy-momentum tensor is traceless.  In two dimensions
this implies that, in light-cone coordinates $(z,\bar z)$, the
component
$T \equiv T_{zz}$ satisfies
\be
\bar\partial T = 0,
\en
and so any polynomial in $T$ is conserved.  Thus the integrals of such
polynomials are conserved charges for the theory.

There is a remarkable connection between
the integrable systems referred to above and the
infinite-dimensional conformal algebra and its
extensions.  Of course any conformally invariant theory in two dimensions
automatically possesses an infinite number of conserved quantities, as
argued above, but
in order to have integrability it is necessary to
select from the set of all polynomials in $T$ a subset for which the
Poisson bracket between any pair of charges is zero. It turns out that
the charges
arrived at in this way are essentially the conserved charges of the
sinh-Gordon and KdV equations, even though these systems are not
conformally invariant.  A similar result holds for the generalizations of
these systems corresponding to arbitrary Lie algebras. For each Lie algebra
there is an extension of the conformal algebra, referred to generically as
a \W-algebra, and choosing a commuting subset of conserved quantities in
the \W-algebra corresponding to the Lie algebra \goth{g} gives the
conserved charges for the integrable systems corresponding to \goth{g}.

Let us spell out in more detail how this procedure works in the simplest
case.  The first step is to write the KdV equation in Hamiltonian form. With
respect to the so-called second Poisson bracket structure for the field
$u(x)$,
\be
\{ u(x), u(y) \} = \left( u(x) \partial_x + \partial_x u(x) - \partial_x^3
\right)\delta(x-y)
\en
the time evolution is given by the Hamiltonian
\be
H_2 = {1\over 16} \int u^2.
\en
It is through this Poisson bracket that the connection with the
conformal algebra is made.  It was recognized in \cite{G} that the
Poisson bracket for the Fourier modes of $u$ is precisely the
two-dimensional conformal algebra, so that we can identify $u$ with the
energy-momentum tensor $T$.  The conserved quantities of the KdV equation
are then integrals of polynomials in $T$ and its derivatives.

If now we take a realization of the second Poisson bracket structure of the
KdV equation in terms of some field $\Phi$, so that $u$ is expressed in
terms of $\Phi$ and the Poisson bracket for $u$ follows from that of
$\Phi$, we can obtain an integrable hierarchy of evolution equations for
$\Phi$.  This hierarchy is obtained simply by taking the conserved charges
for the KdV equation, expressing them in terms of $\Phi$, and then using
these as Hamiltonians to generate time evolutions for $\Phi$. Every
possible realization of the Virasoro algebra leads to such a
hierarchy.

An example of this is the  realization of the second Poisson bracket
structure of the KdV equation given by the Miura transformation.  The field
$u(x)$ is expressed in terms of a field $v(x)$ satisfying
\be
\{ v(x), v(y) \} = \partial_x \delta (x-y)
\en
through the transformation
\be
u={1\over2}v^2 + v'.
\en
The hierarchy of evolution equations obtained in this way for $v(x)$ is
precisely the mKdV hierarchy.  We could, in addition, take
$v$ to be the derivative of a scalar field $\phi$.
The Poisson bracket of $u$ with $\int dx \exp(-\phi(x))$ is then
easily seen to be
zero.  In fact the conserved quantities of the KdV equation, when expressed
in terms of $v$ through the Miura transformation, are even functions of
$v$, and so these charges must also commute with $\int dx \exp(\phi(x))$
and
hence with $\int dx \cosh \phi(x)$.  But $\int dx \cosh \phi(x) $ is the
Hamiltonian for
the sinh-Gordon equation, and so the conserved charges for the KdV equation
are also conserved for the sinh-Gordon equation.  Hence the KdV, mKdV and
sinh-Gordon equations have essentially the same conserved quantities
because their fields provide different realizations of the classical
Virasoro algebra.

There are analogues of this picture for other integrable systems.  In
particular let us consider the $N$'th KdV hierarchy, for variables
$u_i, i = 1,\ldots N,$ corresponding to the Lie algebra $sl(N+1)$.  The
equations of this hierarchy can be written in Hamiltonian form with Poisson
brackets that are a classical limit of the $\W_{(N+1)}$ algebra of Fateev
and Lukyanov \cite{FL}.  There is a realization of the $u_i$ in terms of
fields $\phi_j$ with Poisson brackets $\{\phi_i'(x), \phi_j'(y)\} =
\delta'(x-y)$ via the generalized Miura transformation
\be
\prod_i (\partial + h_i\cdot\phi) = \sum u_s \partial^{N-s},
\en
where the $h_i$ are the weights of the fundamental representation of
$sl(N+1)$.  The evolution equations for the $\phi_i$ generated by the
Hamiltonians of the $N$'th KdV hierarchy are those of the $N$'th mKdV
hierarchy, and these equations have the same conserved quantities as the
affine Toda field theory based on $sl(N+1)$.

The above considerations have all been classical, and it is natural to
consider whether the same picture holds after quantization.  The quantum
KdV, quantum mKdV and quantum sinh-Gordon equations were considered in
\cite{SY}, and the first few commuting conserved quantities were
constructed.  It was indeed found that for these quantities the connection
spelt out above held also at the quantum level. Thus the charges for the
quantum KdV equation, when expressed in terms of a scalar field through the
Feigin-Fuchs construction, were precisely the charges found for the quantum
mKdV and sinh-Gordon equations.  The first few conserved charges for these
systems
have also been found from the perspective of perturbations of conformal
field theories \cite{Z,EY1,HM}.  The existence of conserved commuting
charges at all expected levels has argued in a free-field realization in
\cite{FF}, but the methods that are used classically to construct the
conserved charges and establish their properties have not so far been
carried over to the quantum theories.  The difficulties in doing this can
be traced to the necessity of normal-ordering the quantum conserved
currents and their products.

One approach to establishing the integrability of quantum theories would be
to attempt to exploit the remarkable connection spelt out above at the
classical level between infinite-dimensional algebras and integrability.
Indeed one might conjecture that this pattern is the general case, and that
every integrable system is associated with the Virasoro algebra, or an
extension of it, in this way.  Then different realizations of a given
algebra will lead to different integrable systems, but all such systems
originating from a given algebra will have the same conserved quantities.

Clearly, given an infinite set of commuting quantities, we can define an
integrable system by taking them as Hamiltonians.  In this sense one
can regard the infinite set of commuting quantities as the primary object.
{}From this point of view the conjecture is that any such set of quantities
is associated with the Virasoro algebra or a \W-algebra.  One can also
go further and conjecture that the commuting quantities can be
seen as a Cartan subalgebra of an even larger algebra.

In recent papers we have attempted to shed some light on these issues, and
here we review these developments.  In section 2 we examine commuting
charges that can be constructed from the generators
of a \mbox{\sf U}$(1)$ Kac-Moody algebra. It was found
that there exists an infinite number of series of commuting charges, with each
series itself containing an infinite number of charges.  In terms of the free
field realization of the Virasoro algebra each series exists only for a
particular value of $c$.  The $n$'th series contains even spin currents for
every even value of the spin, but in addition contains odd spin currents at
spins $m(2n-2)+1$, for $m$ a positive integer.  The lowest such non-trivial odd
spin is $2n-1$, and the corresponding current can be taken to be a primary
field. Furthermore this field and the identity operator generate a \W-algebra,
denoted by \W$(2n-1)$, which plays an important role in determining the
structure of the charges. We were able to prove analytically that these charges
did commute, and furthermore we found that the \W-algebra and an extension of
it provide a natural algebraic framework in which this result can be
understood.

As explained above, given an infinite set of commuting currents
we can use them as Hamiltonians to specify a dynamical system
which has the commuting quantities as conserved quatities.
We explain at the end of section 2 how this identification allows
us to conclude that the above sets of commuting currents are the
conserved quantities for Sine-gordon theory for particular values of
the coupling.

A given commuting set does not, however, generate the exceptional
\W-algebra associated with it, since it contains only particular
moments of the generators of this algebra. It is interesting to
ask what theory does possess the exceptional algebras as symmetries.
Liouville theory is well known to be conformally invariant for any
value of the coupling, and in section 3 we show that for specific
values of the coupling this theory admits an extended symmetry,
namely one of the exceptional \W-algebras.  This is a new result which is not
contained in our earlier papers.

In section 4, we review the work of reference \cite{FW}, which asked
whether {\it any} integrable
system can be related to some infinite-dimensional algebra. To
this end we examined the \nls equation.  We will show that there is a
relation between this equation and the KP hierarchy which is analogous to
that which exists between the mKdV and KdV hierarchies.  The essence of the
connection between the \nls equation and the KP hierarchy is that the first
Poisson bracket structure \cite{Wa} of the KP hierarchy is isomorphic
\cite{Ya,YW} to the algebra \Wi, which is a linear extension of the
Virasoro algebra containing a single spin-$i$ quasiprimary field for each
spin $i\ge 1$ \cite{PRS}, and this algebra has a realization in terms of a
complex scalar field $\psi,\psi^*$.
The quantum KP hierarchy is then defined using the quantum \Wi\ algebra,
and
it is shown to possess conserved quantities at least at the first few
levels.  The connection between the classical \nls equation and the KP
hierarchy spelt out above is shown to extend to the quantum analogues.

\sect{Commuting quantities and \W-algebras}
In this section we attempt to gain some insight into the nature of
integrable systems by examining whether we can construct sets of
commuting currents from the simplest possible algebra namely a
$U(1)$ Kac-Moody algebra. We will not assume
in our search that the currents should contain any other algebra.
The \mbox{\sf U}$(1)$ Kac-Moody algebra $\widehat{\mbox{\sf U}(1)}$,
\be
[J_n,J_m] = n \delta_{n+m,0},
\label{KM}
\en
has generators $J_n = \oint dz\, z^n J(z)$ with a
corresponding operator product expansion
\be
J(z) J(w) = {1 \over (z-w)^2}+ \ldots\quad.
\label{JJ}
\en
To be precise we wish to address the question of what sets of mutually
commuting quantities can be constructed as integrals of
polynomials in $J$ and its derivatives.  To put this another
way, we want to find sets of mutually commuting operators in
the enveloping algebra of $\widehat {\mbox{\sf U}(1)}$. We know that it
is possible to write a stress energy tensor in terms of $J$,
\be
T=\sm{1}{2} :J^2: + \alpha J',
\label{FF}
\en
and so we certainly expect to find the commuting charges
that correspond to the KdV equation.  We
shall therefore look for sets of mutually commuting charges
that include operators not contained in this series of charges.

Given currents $A$ and $B$ with charges
\be
Q_A=\oint dz\,A(z),\quad Q_B = \oint dz\,B(z),
\en
the commutator of $Q_A$ with $Q_B$ is given by
\be
[Q_A,Q_B] = \oint_0 dw \oint_w dz\, A(z) B(w).
\label{commutator}
\en
The only contribution to this commutator comes from the
single pole term in the OPE of $A$ and $B$, so
$Q_A$ and $Q_B$ will commute precisely when this single pole
term is a derivative.

It is straightforward to see that
\be
Q_1 \equiv \oint dz\, J(z)
\label{Q_1}
\en
commutes with any charge constructed from $J$ and its
derivatives.  It is also true that
\be
Q_2 \equiv \oint dz\, :\!J^{\,2} (z)\!:
\label{Q_2}
\en
commutes with all other charges, since if $P$ is an
arbitrary differential polynomial in $J$, the single pole
term in $:\!J^2(z)\!: P(w)$ comes from the single contraction
term, namely
\be
2 \sum_{n=0}^\infty{(n+1)! \over (z-w)^{n+2}} :\!J(z)\,
{\partial P(w) \over \partial J^{\,(n)}(w)}\!:\ ,
\en
and the coefficient of the single pole in this expression is
just
\be
2 \sum_{n=0}^\infty :\!J^{\,(n+1)}\, {\partial P \over \partial
J^{\,(n)}}\!:\ \  =2 P'
\en
This is also apparent from the Feigin-Fuchs
representation---the integral of $1/2 :\!\!J^{\,2}\!\!:$ is just $L_{-1}$,
the generator of translations.

In order to go beyond these somewhat trivial commuting
charges, we must specify some rules for the game;
we will look for sets of  charges commuting with the
integral $Q_4$ of the spin-4 current
\be
p_4 = \,:\!J^{\,4}\!: +\, g :\!J'^{\,2}\!:
\label{p_4}
\en
for some value of the coupling $g$, with the same value of $g$ for
each current in a given set.  Moreover, we require that each
set should contain a current of odd spin.
We found that there were a number of different sets of such
charges:

{{ \bf  There exists an infinite number of sets of
mutually commuting operators constructed from the Kac-Moody
generators $J(z)$.  The series $S_n$ has a unique current at
every even spin and unique odd spin currents at spins
$1+m (h-1) $ for $h=2n-1$ and $m=0,1,2,\ldots$.}}

Thus the first series $S_2$ has $h=3$ and
has a unique current at every odd spin as well as at every
even spin.  The second series $S_3$ has unique odd spin
currents at spins $5,9,13,\ldots$ as well as currents at
every even spin.

We found such sets of odd spin currents for every odd
integer $h$ larger than $1$. In fact each set was
uniquely fixed by the spin of the
lowest non-trivial odd spin current.  The
charges constructed from these currents commute with $Q_4$
provided that $g$ is given in terms of $h$ by
\be
g={1\over h+1} (h^2-4h-1)
\label{g,h}
\en

Given a set of charges commuting with a given spin-3 charge
$Q_4$, it is natural to ask whether these charges commute
with each other.  We have verified that this is indeed the
case.

The series of currents described above were found initially by using
Mathematica \cite{W}.  It is a remarkable fact that there is a relatively
simple explicit formula that gives all the currents in all the series, at
least up to the dimension to which we calculated, namely spin 13.
We refer the reader to our paper \cite{FHW} for details of
this formula, since we will not need them here.

We can now ask whether we can find some underlying  algebraic
structure that would explain the above set of currents. The first step in this
process is the verification that the even spin currents can
be expressed in terms of the Feigin-Fuchs energy-momentum
tensor given in eqn (\ref{FF}), which has background charge
$\alpha$ and central charge $c= 1-12 \alpha^2$.
Evidently $Q_2 = \oint \!:\!J^{\,2}\!: \,= 2 L_{-1}$.  The charge $Q_4$
can also be written in terms of $T(z)$, provided we relate
the coupling $g$ and the central charge $c$ by
\be
g=-\sm{1}{3} (c+5);
\label{g,c}
\en
for this value of $g$ we have
\bea
Q_4 & \equiv & \oint :\!J^{\,4}\!: +\, g\, :\!J'^{\,2}\!:\nonumber \\
    & = & 4 \oint :\!T^2\!:\quad.
\label{TJ}
\ena
Here $T^2$ is normal-ordered according to the standard prescription in
conformal field theory, which takes the normal product of two
operators $A(z)$ and $B(w)$ to be the term of order $(z-w)^0$ in their
OPE \cite{BBSS},
\be
(AB)(Z) \equiv \oint \, dw { A(w) B(z) \over w-z }.
\label{NOP}
\en
To reconcile the two expressions for $Q_4$ in (\ref{TJ}), we must relate
the normal
ordering in terms of $J$, which is the usual Wick
ordering, to that in terms of $T$.  Using the formula \cite{BBSS}
\be
((AB)C) = A(BC) + [(AB),C] - A[B,C] - [A,C]B
\en
with $A=B=J$ and $C=\,:\!J^2\!:$, and also making use of
\be
:\!J:\!J^2\!:: - ::\!J^2\!:J \!:\,= -J'',
\en
it is straightforward to recover eqn(\ref{TJ}).

In this way it is possible to express the even spin currents in
terms of $T$ alone, provided we use the freedom to add and
subtract total derivatives, but the odd spin currents cannot be expressed only
in terms of $T$. To summarize the result:

{{ \bf  The even spin currents can be built in terms of
the energy-momentum tensor $T={1\over 2} J^{\,2} + 3{(h-1)\over \sqrt{h+1}}
J'$ where the central charge is $c = {1\over h+1}(-3h^2 +
7h -2)$.}}

The value $c_n$ of the central charge corresponding to the
series $S_n$ can be written in the form
\be
c_n = 13 -6n -{6\over n}.
\label{cn}
\en

We now consider if some even larger algebra can play a role.
It is known for $n=2$, 3 and 4 that a \W-algebra \W$(2n-1)$ exists \cite{KW,B},
where \W$(2n-1)$ is an algebra generated by the identity operator together with
a single primary field of spin $2n-1$.  For $n=2$ this algebra is the
\W$(3)$-algebra of Zamolodchikov \cite{Z2}, which exists for any value of $c$,
while for $n=3$ the algebra \W$(5)$ exists for only five values of $c$, namely
6/7, $-350/11$, $-7$, $134\pm 60/\sqrt5$, and for $n=4$ \W$(7)$ exists for the
single value $-25/2$ of $c$.  There is an argument to suggest that an algebra
\W$(2n-1)$ exists for any $n$ at the value of $c$ given in eqn (\ref{cn}).
This can be seen by examining the primary field algebra for the conformal field
theory with $c$ given by (\ref{cn}).  In this model the field $\phi_{(3,1)}$
has dimension $2n-1$ and is expected to have an OPE with itself of the form
\cite{BPZ}
\be
[\phi_{(3,1)}][\phi_{(3,1)}] = [1] + [\phi_{(3,1)}] + [\phi_{(5,1)}].
\label{3,1,3,1}
\en
The dimension of $[\phi_{(5,1)}]$ is $6n -2$, however, which is too high to
appear in the OPE of a spin $2n-1$ field with itself, and so we expect the
identity operator together with $[\phi_{(3,1)}]$ to form a closed operator
algebra.   This observation has also been made previously by Kausch \cite{K}.
We note in addition that for the values of $c$ we are considering the field
$[\phi_{(3,1)}]$ has odd spin, and as a consequence cannot appear on the
right-hand side of the OPE (\ref{3,1,3,1}).  Thus the OPE of $[\phi_{(3,1)}]$
with itself gives only the identity operator in this case.

It is natural to
suppose that the series $S_n$ is associated in some way with
the algebra \W$(2n-1)$.  This is confirmed by the realization that
the first non-trivial field of odd spin in $S_n$ can be
chosen to be a primary field and that its OPE with itself is
just that for \W$(2n-1)$.

We have amassed sufficient evidence to conclude that:

{{ \bf  The series $S_n$ can be associated with the
algebra \W$(2n-1)$, the current $p_{2n-1}$ being a primary field which
is the generator of \W$(2n-1)$.  All odd spin currents in $S_n$ are descendants
of $p_{2n-1}$.}}

All the above deliberations were based on
computer results using the Mathematica programme of Thielemans \cite{T}.
The calculations involved  currents of spin 13 and less.

One might ask how the above sets of commuting quantities compare with those
that are known from classical integrable systems.  For the $n$'th KdV hierarchy
there are conserved currents with spins 2, 3, $\ldots$, $n+1$, mod $n+1$, so
that the conserved charges have spins 1, 2, $\ldots$, n mod $n+1$.  These are
the exponents of the Lie algebra $sl(n)$ repeated modulo the Coxeter number,
and in fact for any Lie algebra $\goth g$ there exists an integrable system for
which the charges have spins equal to the exponents of $\goth g$ \cite{Wi,DS}.
These do not correspond to the spins of the currents for any of the series
found above. The even spins however are those found for the quantum KdV system
since there exists only one set of commuting even spins for a
given central charge.

We now study in more detail the first series of currents,
$S_2$, containing a current $J^{\,3}$.  Some features of this series will
generalize to the other series, although there are aspects of this
series that will have no analogues for the others.
Let us first show that this series, which is defined by the existence of
commuting currents\footnote{We will loosely say that two currents commute
if their corresponding charges commute} of spins 3 and 4, corresponds to
the central charge having the value $c=-2$.  If we take an arbitrary
current $p_\ii$ of spin $\ii$, defined modulo derivatives,
\be
p_\ii = J^{\,\ii} + g(\ii) J^{\,\ii-4}(J')^2 + \ldots,
\en
it is straightforward to calculate its OPE with $J^{\,3}$.  We find that,
up to
derivatives, the coefficient of the single pole is given by
\be
{\ii(\ii-1)(\ii-2)(\ii-3)\over 4}J^{\,\ii-4} (J')^3 + 6 g(\ii)
J^{\,\ii-4}(J')^3+\ldots,
\en
so we must have
\be
g(\ii) = - {\ii \choose 4}.
\en
For $\ii=4$ we find the current is $J^{\,4}-(J')^2$, and comparing with
(\ref{TJ}) we find that $c=-2$.  Hence we conclude that a spin-3 and a
spin-4 current can commute only if $c=-2$.

One recognizes from the computer results of reference \cite{FHW} that, for
$c=-2$, there is a considerable simplification in the form of the
currents.  In fact they are consistent with the formula
\be
p_\ii = \,:\!e^{-\phi}\partial^\ii e^\phi\!:\quad,
\label{p_\ii,c=-2}
\en
up to derivatives.
The generating function for this series is
\be
:\!e^{-\phi(z)+\phi(z+\alpha)}\!:\,=\sum_{r=0}^\infty {\alpha^r\over
r!}p_r(z).
\en
It is straight-forward to show that the integrals of the
currents $p_\ii$ commute by considering the OPE of the above exponential
with itself.

These results for the series $S_2$ can understood in terms of
the algebra \W$_\infty$ of reference \cite{PRS1,PRS2}.  This is a linear
algebra containing a quasiprimary field $V^i(z)$ with spin $i+2$ for
$i=0,1,2,\ldots$.  Examining the commutation relations of \W$_\infty$, as
given for example in eqn(3.2) of \cite{P}, it can be seen that the modes
$V^i_{-i-1}$ form an infinite set of commuting operators, which we might
think of as a Cartan subalgebra.  Since $V^0(z) = T(z)$, we recognize
$V^0_{-1}$ as $L_{-1}$, and similarly $V^1_{-2}$ is the mode $W_{-2}$ of
the spin-3 primary field $W(z)$.

For the case of $c=-2$, however, the $\W_\infty$ algebra has a realization
in terms of a complex fermion, and this can be bosonized to give a
realization in terms of a single real scalar field.  In terms of this
scalar field $\phi$, the field $V^i$ is proportional to
$:\!e^{-\phi}\partial^{i+2}e^\phi\!:$, up to derivatives.  We recognize
this as
the current given in eqn (\ref{p_\ii,c=-2}).  We also note that for
$c=-2$ the fields $V^i$ for $i\ge2$ can be written as composites of the
stress tensor $T$ and the spin-3 primary field $W$, so that the enveloping
algebra of the Zamolodchikov \W$(3)$-algebra
contains \W$_\infty$ as a linear subalgebra for $c=-2$.  This thus gives a
nice explanation of the presence of the infinitely many commuting
quantities in the enveloping algebra of \W$(3)$ in terms of a linear
algebra.

Let us summarize the main results for the series of currents $S_3$.
Starting from a \mbox{\sf U}$(1)$
Kac-Moody algebra, we demanded a set of commuting currents that included
currents of spins 3 and 4.  We found that we could supplement these two
currents by an infinite number of other currents, one at each spin, that
formed a mutually commuting set.  These could be expressed in terms of a
stress tensor $T$ and a spin-3 primary field $W$ alone, where $T$ has a
central charge $c=-2$.  Furthermore these currents can be identified with a
Cartan subalgebra of \W$_\infty$, which in turn can be written in terms of
$T$ and $W$.  We therefore see that in terms of the algebra \W$_\infty$ it is
possible to understand very easily the existence of the commuting charges.
Although we do not know of a generalization of \W$_\infty$ that enables us to
gain a similar understanding of all the series of commuting charges, we shall
see below that, for each series, there is a relation between the
odd-spin currents that allows to us to give a simple proof of the commutativity
of the corresponding charges.

We shall now give explicit
formulae for all of the odd spin fields in each of the series $S_n$, and we
shall prove that these fields commute with each other and with all of the
even spin fields in the series\footnote{Here, as before, we refer to fields
commuting when we really mean that their integrals commute.}.  We shall
also spell out in detail the connection of the series $S_n$ to a
\W-algebra.

Let us consider the forms of the fields of odd spin in $S_n$.  The series
$S_n$ exists for $c$ having the value $13-6n-
6/n$, with $n$ a positive integer, and the first
non-trivial field of odd spin has dimension $2n-1$ and can be
chosen to be primary.  From the work of Zamolodchikov \cite{Z} we expect there
to be
only two primary fields that commute with the even spin fields built out of
the stress tensor, namely the two fields with null descendants at level 3.
If we write an arbitrary value of the central charge as $c=13-6t-6/t$, the
two corresponding fields will have dimensions $h_{(3,1)}=2t-1$ and
$h_{(1,3)}=2/t-1$.  These two values of the dimension $h$ are related to
$c$ by
\bea
c&=&13-3(h+1)-{12\over h+1}\nonumber\\
&=& {1\over h+1}(-3h^2+7h-2).
\ena
Taking $t$ to be a positive integer $n$ we indeed expect to find a field of
dimension $2n-1$ that commutes with the commuting charges constructed from
$T$.  In order to write down an explicit
expression for this field in terms of the currents $J$
alone, we first consider a field of this dimension that can be
written as an exponential of $\phi$, namely
\be
V_{-\alpha_+} \equiv e^{ -\alpha_+ \phi}
\en

where $\alpha_+ = \sqrt{2n}$ and $\alpha_- = -\sqrt{2/n}$.
There exist one other field with the same weight, but we shall concentrate our
attention on
the above field in what follows.

When $c=13-6n-6/n$, we can use the screening charge
$Q_+ \equiv \oint \exp \alpha_+ \phi$ to construct another primary field of
weight $2n-1$ that is annihilated by ${\cal O}_{(3,1)}$.  The commutator of
a Virasoro generator $L_n$ with $\exp \alpha_+ \phi$ is a total derivative,
so provided this field is single-valued the commutator of $Q_+$ with a
primary field will be another primary field of the same weight.  In general
this new primary
field could be zero, but that will not be the case here.  For the
values of $c$ in which we are interested, $\exp \alpha_+ \phi$ is indeed
local with respect to $\exp - \alpha_+ \phi$, and we can therefore
construct
another primary field of weight $2n-1$ by
\be
p_{2n-1}\equiv [Q_+,\exp - \alpha_+ \phi(z)] =
\oint_z dw \exp(\alpha_+ \phi(w))\exp(-\alpha_+ \phi(z)).
\en
This field has vanishing background charge and so is expressible entirely
in terms of the current $J$ and its
derivatives, and in fact it is given explicitly by
the term of order $\partial^0$ in the differential operator
\be
(\partial + \alpha_+ J)^{2n-1},
\en
in which the derivatives are taken to act on everything that occurs to the
right of them.
We claim that this is the first non-trivial field of odd spin occurring in
$S_n$. For the first few series this can be checked explicitly against the
computer results given in Section 2. To show that it is true in general, we
need to check that the charge constructed from this field
commutes with all of the odd spin charges obtained by integrating
the even spin currents, which correspond to those that
occur in the quantum KdV equation. This
is a consequence of the fact that the screening charge
$\exp \alpha_+ \phi$ commutes with
any polynomial in $T$ and its derivatives, and in particular with the odd spin
charges.  Since these odd spin charges are even under
$\phi \rightarrow - \phi$, it follows that $\oint \exp - \alpha_+\phi$ also
commutes with these charges, as explained in ref \cite{SY}.

The above considerations immediately suggest a way to
construct infinitely many odd spin currents whose integrals  commute with the
odd spin charges.
We take $Q_+$, as defined earlier, to be $\oint \exp \alpha_+ \phi$, and we
define also $Q_- = \oint \exp - \alpha_+ \phi$.  These two charges have
conformal weights $0$ and $2n-2$ respectively, and each commutes with all
of the odd spin charges.
We then define an operator $\Delta$ that acts on an arbitrary field
$\Phi$ by
\bea
\Delta \Phi(z) &=& [Q_+,[Q_-,\Phi(z)]]\nonumber\\
&=& \oint_z dy \oint_{y,z} dx :\!e^{\alpha_+ \phi(x)}\!:\,:\!e^{-\alpha_+
\phi(y)}\!: \Phi(z).
\ena
If the integral of $\Phi$ commutes with the odd spin charges, it is clear that
the integral of $\Delta \Phi$ will also commute with these charges.
Our  strategy is then to apply $\Delta$ repeatedly to the primary field
constructed above.  In fact it is interesting to note that the primary
field $p_{2n-1}$ can itself be written as
\be
p_{2n-1} = \Delta J,
\en
since $[Q_-, J(z)]=\exp-\alpha_+\phi(z)$.
If we write $\Phi^{(m)}$ for the $m$'th odd spin field in $S_n$, and $Q^{(m)}$
for the corresponding charge, we have
\be
\Phi^{(m)} = \Delta \Phi^{(m-1)},\quad \Phi^{(0)}=J,
\en
with similar formulae holding for $Q^{(m)}$.
Thus $\Phi^{(m)}$ is given by the multiple commutator
\be
\Phi^{(m)} = \Delta^m J =
\underbrace{[Q_+,[Q_-,\ldots[Q_+,[Q_-}_{m\rm\;times},J].
\en
Furthermore it is readily
seen to be expressible as a sum of terms, with each term being a product of
derivatives of exponentials of $\phi$, with the total \mbox{\sf U}$(1)$ charge
for each term in the sum being zero.  Thus $\Phi^{(m)}$ can be written in terms
of $J$ alone, with no exponentials of
$\phi$.  While it is conceivable that this multiple commutator could vanish,
and indeed this will be the case for some orderings of the $Q_+$ and
$Q_-$, we believe this will not happen in general for the ordering we have
chosen.  In fact  for the first series, it is straightforward to show that the
action of $\Delta $ does not vanish and that it ladders
one up the odd  currents of the series.

Having seen how to write explicit formulae for an infinite set of charges
commuting with the even spin charges, it is necessary to understand why these
charges commute amongst themselves.
The details are given in reference\cite{FHW}, and we omit them here.
The basic strategy, however, is to use the crucial identities
\be
[Q_+,[Q_+,[Q_+,Q_-]]] = 0,
\label{Q_+^3}
\en
and
\be
[Q_-,[Q_-,[Q_-,Q_+]]] = 0,
\label{Q_-^3}
\en
which can be obtained very simply just from looking at the
contour integrals that need to be done in order to evaluate these
expressions.
Making use of these relations one can then prove by induction that
\bea
[Q^{(i)},Q^{(j)}] & = & 0, \qquad\hbox{for}\quad i + j = n,\nonumber\\[1mm]
Q^{(n)} & \equiv & Q_+Q_-Q^{(n-1)}\nonumber\\[1mm]
        & = & [Q_-Q^{(k)},Q_+Q^{(l)}], \qquad\hbox{for}\quad k+l=n-1,
\nonumber\\[1mm]
Q_+^2Q^{(n-1)} & = & Q_-^2Q^{(n-1)} = 0.
\label{hypothesis}
\ena
In particular we have that $[Q^{(i)},Q^{(j)}]=0$ for all $i$ and $j$.

It is interesting to ask to what perturbation of a conformal field theory with
$c=13-6n-6/n$ the commuting charges found in this paper correspond.  This
amounts to finding an operator that commutes with these charges.  The charges
constructed from the even spin currents are differential polynomials in $T$ and
so will commute with the screening charges $\oint\exp\alpha_\pm\phi$. They also
commute with $\oint\exp-\alpha_\pm\phi$, on account of being even polynomials
in $\phi$.  The charges for the odd spin currents, however, are not constructed
from $T$ alone, and so will not commute with the above charges in general.
Nevertheless, since they are constructed from $\oint\exp\alpha_+\phi$ and
$\oint-\exp\alpha_+\phi$, and since $\alpha_+\alpha_- = -2$ implies that both
of these charges commute with $\oint\exp\pm\alpha_-\phi$, all the charges we
have found commute with $\oint\exp\alpha_-\phi$ and $\oint\exp-\alpha_-\phi$.
While the former is the screening charge, the latter is the field
$\phi_{(1,3)}$, which has weight $2/n-1$.  Thus the commuting charges are
conserved in the presence of the $\phi_{(1,3)}$ perturbation.  These remarks
are in agreement with those of Eguchi and Yang \cite{EY2}, who considered the
quantum sine-Gordon theory with Hamiltonian $\oint e^{\alpha_-\phi} -
e^{-\alpha_-\phi}$.  They observed that for particular values of the
sine-Gordon coupling constant extra odd-spin conserved currents existed for
spins $2n-1$ mod $(2n-2)$.  The detailed forms of their charges are different
from ours,
however, and it is not possible to write them in terms of $J$ alone. Their
first
charge, for example, is given in our notation by $Q_+-Q_-$.

We now turn to the connection between the series of commuting charges and
certain \W-algebras.  We explained in the previous section that the first
series of charges was related to Zamolodchikov's \W$(3)$-algebra with the
central charge taking the value $c=-2$, and that for this value of $c$ the
enveloping algebra of the \W$(3)$-algebra contains a linear subalgebra which
is just \W$_\infty$.  In order to make a connection between \W-algebras and
the higher series, there was another aspect of the algebras that played an
important role in our proof of the commutativity of the charges constructed
from the odd-spin currents.  This was the fact that we had not just a
single primary field of conformal weight $2n-1$ for $c=13-6n-6/n$, but in
fact we made use of three distinct primary fields having this dimension.
In our free-field representation, these were $\exp (-\alpha_+ \phi)$, $Q_+
\exp (-\alpha_+ \phi)$ and $Q_+^2 \exp (-\alpha_+ \phi)$.  It is known that
for the values of $c$ that we are considering there exist \W-algebras
generated by the stress tensor and three primary fields each having spin
$2n-1$ \cite{K}. Let us denote these algebras by \W$((2n-1)^3)$. If we denote
the primary fields by $W^i(z)$, for $i=1$,2 and 3, they have operator
product expansions of the form
\be
[W^i][W^j] = \delta^{ij}[I] + \epsilon^{ijk}[W^k],
\en
so that there is an \mbox{\sf SU}$(2)$-like structure present.  The operator
$Q_+$ can be considered as an \mbox{\sf SU}$(2)$ raising operator.  We have
seen that acting repeatedly with $Q_-$ on the multiplet of primary fields gives
other spin-1 \mbox{\sf SU}$(2)$ multiplets of higher conformal weight.  The
fields in these higher multiplets are no longer primary, but their integrals
give rise to the infinite sets of commuting charges we have found.

\sect{Louiville theory and the exceptional \W-algebras }

We have found that  a given set of commuting currents ,$S_n$,
was a symmetry of the sine-Gordon theory for a
particular value of the coupling. This theory, however, does not
possess the full exceptional \W-symmetry since the
Hamiltonian only commutes with a specific
mode of the \W-current and the higher currents. A natural
question to ask is what theory does possess the exceptional
\W-algebra as a symmetry. The $\W(2n-1)$ current $J(z)$ is of
the form
$$ j(z) =[Q_{+},e^{\alpha_{+} \phi(z)}].$$
Let us consider the Hamiltonian
$$H=\oint dz:\oint dz\,e^{\alpha_{-} \phi(z) }:.$$
Since $H$ is a screening charge, it commutes with the Virasoro
generators $L_n$ and also with $Q_{+}$, since $\alpha_{+} \alpha_{-}=-2$
and
$$[H,Q_{+}]=- \oint_0 dz \oint_z dw e^{\alpha_{-} \phi(z) }e^{\alpha_{+}
\phi(w)}$$
 $$=-\oint_0 dz \oint_z dw (z-w)^{-2}e^{\alpha_{-} \phi(z) +\alpha_{+}
\phi(w)}$$
 $$=\oint_0 dz \alpha_{+} \partial \phi e^{(\alpha_{-} + \alpha_{+}
)\phi(z)}=0$$
Similarly, $H$ commutes with $ e^{-\alpha_{+} \phi(z)}$ and consequently
$Q_{-}$. It follows therefore that $H$ commutes with all the modes of the
current $j(z)$, and so the  exceptional $\W(2n-1)$ algebra which we conclude is
a symmetry of Louville theory for the above coupling.

It also follows that all the moments of the higher currents are
also a symmetry of this Louville theory since
$$[Q_{+},[Q_{-},......[Q_{+},e^{-\alpha_{-} \phi(z)}]...]].$$
also commutes with $H$.

\sect{\Wi\ and the \nls equation}

We now turn our attention to the \nls equation, which is well-known
to be integrable, in an attempt to understand the connection of
this to a \W-algebra.  We do this by first establishing a
connection of the \nls equation to the KP hierarchy, which is in
turn known to be related to the algebra \Wi.  We then extend these
considerations to the quantum domain.

The \nls equation is given by
\be
{\partial\psi\over\partial t} + \psi'' + 2 \kappa\, \psi^* \psi^2 = 0,
\en
where $\psi,\psi^*$ is a bosonic complex scalar field and $\kappa$ is a
coupling
constant.  This equation can be written in Hamiltonian form
$\dot\psi=\{\psi,H\}$ provided we adopt the Poisson brackets
\be
\{\psi^*(x),\psi(y)\} = \delta(x-y),
\quad\{\psi(x),\psi(y)\} = \{\psi^*(x),\psi^*(y)\}=0
\label{PBnls}
\en
and take $H=\int(\psi^*\psi''+ \kappa\,\psi^{*2}\psi^2)$.

It is useful to note that it is possible to assign weights to $\psi$ and
$\psi^*$ in a manner consistent with the Poisson bracket.  We take
$\partial$ and $\delta(x-y)$ to have weight 1, and  then eqn (\ref{PBnls})
implies that the product $\psi^*\psi$ must have weight 1.  If we now take
$\kappa$ to have weight 1, the Hamiltonian $H$ will be homogeneous of
weight 2.

The \nls equation possesses an infinite set of conserved quantities $Q_n$,
given by the formula
\be
Q_n = {1\over \kappa}\int dx\,\psi(x) Y_n(\psi,\psi^*),
\en
where
\be
Y_1 = \kappa\, \psi^*,\qquad
Y_{n+1}=Y_n'+\psi \sum_{k=1}^{n-1} Y_k Y_{n-k}\quad\hbox{for $n \ge 1$}.
\en
The first few such charges are
\bea
Q_1 &=& \int dx\,\psi^* \psi\nonumber\\
Q_2 &=& - \int dx\, \psi^* \psi'\nonumber\\
Q_3 &=& \int dx\,(\psi^* \psi'' + \kappa\, \psi^{*2} \psi^2);
\ena
it can be checked that these charges commute and so can be used to generate
commuting evolutions, giving a
hierarchy of equations.

The KP hierarchy consists of an infinite set of differential equations
which has, at first sight, no relation to the \nls equation.  This set of
equations is defined in terms of a pseudodifferential operator
$Q$ of the form
\be
Q = D + q_0 D^{-1} + q_1 D^{-2} + \ldots\quad.
\label{Q}
\en
Here $D$ denotes $\partial/\partial z$ and satisfies the following
generalization of the Leibniz rule:
\be
D^n f = \sum_{r=0}^\infty { n \choose r} f^{(r)} D^{n-r},\quad n \in
\bb{Z}.
\en
An infinite set of commuting time evolutions for the fields
$q_i$, $i=0,1,2,\ldots,$ is given by
\be
{\partial Q \over \partial t_p} = [ (Q^p)_+,Q], \quad p=0,1,2,\ldots
\label{KPevolution}
\en
where $(Q^p)_+$ is the part of $Q^p$ involving no negative powers of $D$.
These evolution equations can be written in Hamiltonian form.  There are
in fact a number of ways to do this, but for our purposes it is the
so-called first Hamiltonian structure that is needed.  The Poisson
bracket in this case is obtained from the method of coadjoint orbits
applied to the Lie algebra of differential operators.  We omit the
details here and simply give the results \cite{Wa}:
\be
\{q_i(x),q_j(y)\} = \sum_{r=0}^j {j\choose r} \partial_x{}^r (q_{i+j-r}(x)
\delta(x-y)) -
\sum_{r=0}^i {i \choose r} (-1)^r q_{i+j-r} \partial_x{}^r
\delta(x-y).
\label{PB}
\en
This algebra is in fact precisely \Wi\ with zero
central charge \cite{Ya,YW}; the $q_i$ are related to the usual basis for
\Wi, denoted by $V^i$ \cite{PRS,P}, by a linear
transformation.  It can be shown that the time
evolutions (\ref{KPevolution}) can be written as
\be
{\partial q_i \over \partial t_p} = \{q_i,H_{p+1}\}, \quad p=0,1,2,\ldots,
\en
where
\be
H_p = {1\over p} tr(Q^{p}).
\en
The trace of a pseudodifferential operator is given by the integral
of the coefficient of $\partial^{-1}$.
Just as for the \nls equation, there is an assignment of weights to the
$q_i$ which is consistent with the Poisson bracket structure (\ref{PB}).
If we
take $\partial$ and $\delta(x-y)$ to have weight 1, then $q_i$ must have
weight $i+1$.  With this choice, the operator $Q$ in eqn (\ref{Q}) does not
have a definite weight, but we can correct this by working instead with
\be
\Q \equiv \kappa^{-1} D + q_0 D^{-1} + q_1 D^{-2} + \ldots,
\en
where, as before, $\kappa$ is a constant of weight 1.  \Q\ then has weight
zero.  Using
\Q\ instead of $Q$ does not change the Poisson brackets between the $q_i$,
but the Hamiltonians
\be
\h_p = {1\over p} tr(\Q^{p})
\en
now involve various powers of $\kappa$.  The first few $\h_p$ are given
explicitly by
\bea
\h_1 &=& \int dx\, q_0\nonumber\\
\h_2 &=& \kappa^{-1} \int dx\,q_1\\
\h_3 &=& \kappa^{-2} \int dx\,(q_2 + \kappa q_0{}^2)\nonumber\\
\h_4 &=& \kappa^{-4} \int dx\,(q_3 + 3 \kappa q_0 q_1)\nonumber.
\ena
The $\h_p$ can be viewed as a set of Poisson commuting quantities within
the enveloping algebra of \Wi.

Having given the KP and the \nls equations, we are now in a position to
explain the relation between them by exploiting the fact that \Wi\ has a
realization in terms of a single complex boson.  Let us take a general
approach first.  We consider a realization of \Wi\ in terms of some set of
fields $\phi_i$, so that the $q_i$ can be expressed in terms of the
$\phi_i$ and the Poisson brackets for the $q_i$ follow from those of the
$\phi_i$.  If such a realization of \Wi\ has zero central charge, we can
clearly construct an infinite set of mutually commuting quantities by
expressing the $H_n$'s in terms of the $\phi_i$.  We can then define time
evolutions for the $\phi_i$ by
\be
{\partial \phi_i \over \partial t_p} = \{\phi_i,H_{p+1}\}, \quad
p=0,1,2,\ldots,
\en
and it follows from the Leibniz property of the Poisson bracket that this
implies the KP evolutions for the $q_i$.

There is a centreless realization of \Wi\ in terms of a complex boson with
Poisson brackets given by
\be
\{\psi^*(x),\psi(y)\} = \delta(x-y),
\quad\{\psi(x),\psi(y)\} = \{\psi^*(x),\psi^*(y)\}=0.
\en
This is most easily seen by checking that the correct Poisson bracket
(\ref{PB})  between the $q_i$ is obtained by taking
\be
q_i = (-1)^i \psi^* \psi^{(i)}.
\en
The first few Hamiltonians of the KP hierarchy, when expressed in terms of
$\psi$ and $\psi^*$, take the forms
\bea
\h_1(\psi^*,\psi) &=& \int dx\,\psi^*\psi\nonumber\\
\h_2(\psi^*,\psi) &=& -  \kappa^{-1} \int dx\, \psi^*\psi'\\
\h_3(\psi^*,\psi) &=&  \kappa^{-2} \int dx\,
(\psi*\psi'' + \kappa\,\psi^{*2}\psi^2)\nonumber\\
\h_4(\psi^*,\psi) &=& -  \kappa^{-4} \int dx\,
(\psi^*\psi''' + 3 \kappa\, \psi^{*2}\psi\psi').\nonumber
\ena
We recognize these immediately as the first few conserved charges of the
\nls equation, up to overall factors.  It follows from the commutativity of
the KP Hamiltonians
that the $\h_n(\psi^*,\psi)$ will also commute.  However, since the charges
of the \nls equation are uniquely determined by the requirement that they
commute with $\h_3$, it follows that the charges obtained from the KP
system must coincide with those of the \nls equation at all levels.

Given that the classical \nls equation is related to the KP hierarchy and
that the quantum \nls equation is known to be integrable, one might
expect there to exist an integrable quantum KP hierarchy.  One way to
construct such a
hierarchy would be
to take the first Poisson bracket structure for the KP hierarchy, which is
isomorphic to \Wi, and then to replace Poisson brackets by commutators.
The existence of an integrable quantum
KP hierarchy would then follow if we could find an infinite set of
commuting Hamiltonians in the enveloping algebra of \Wi.

In this section we construct some quantum analogues of the Hamiltonians
$\h_p$ from
the generators $V^i$ of \Wi.  We use the notation of \cite{PRS}, where
$V^i$ is a quasiprimary field of weight $i+2$.  We used a computer
to look for commuting charges that could be expressed as integrals
of sums of products of the fields in \Wi, using the OPEdefs package of
Thielemans \cite{T}.The first two charges, namely
\be
Q_1 = \int dz\, V^{-1}
\en
and
\be
Q_2 = \int dz\,V^0
\en
are trivial, in the sense that the first of these commutes with all
$V^i(z)$ while the commutator of $Q_2$ with $V^i(z)$ is just
$V^i{}'(z)$. For the first non-trivial charge
we take
\be
Q_3 \equiv \int dz\,\left\{V^1 + \kappa\,(V^{-1})^2\right\};
\en
this is analogous to the classical Hamiltonian $H_3$ of the KP hierarchy
in
that it involves terms of weights 2 and 3.  In this and subsequent
expressions involving products of operators, we use the normal ordering
standard in conformal field theory, namely we take
\be
(AB)(z) = \oint_z dw \, {A(w) B(z) \over w-z}.
\en
It is our belief that there are infinitely many further charges that
commute with the charge $Q_3$, and that furthermore these charges are
unique.  We  have found three such charges, and we have verified that
all the charges we have obtained commute amongst themselves.  We
conjecture
that there exists an infinite set of such charges, one for each spin.  We
give here $Q_4$ and $Q_5$:
\bea
Q_4 &=& \int dz\,\left\{V^2 + 3 \kappa\,(V^{-1} V^0) + {3 c\,\kappa^2/2}\>
(V^{-1})^2\right\}\nonumber\\
Q_5 &=& \int dz\,\left\{ V^3 + \kappa\left( 4(V^{-1} V^1) + 2(V^0)^2 -
5/6\> (V^{-1}{}')^2\right)\right.\nonumber\\
&&\left. + \kappa^2 \left( 2(V^{-1})^3 + 4c\,(V^{-1}V^0)\right)
+ 2c^2\,\kappa^3\, (V^{-1})^2\right\}.
\ena

There is thus strong evidence that quantization of the first Poisson
bracket structure of the KP hierarchy leads to an integrable system.
We now relate this to the quantum \nls equation.
The integrability of the quantum \nls equation has been extensively studied
using the quantum inverse scattering method and also from the point of view
of the existence of an infinite number of commuting conserved quantities.
The relation between these different approaches was discussed in
\cite{OSSY}.  In order to quantize the \nls equation we adopt the following
OPE for the bosonic operators $\psi$ and $\psi^*$:
\be
\psi^*(z) \psi(w) = (z-w)^{-1} + \ldots\quad.
\label{ope}
\en
The time evolution is generated by
\be
H = \oint dz \, (:\psi^*\psi'': + :\kappa \psi^{*2} \psi^2:),
\en
in the sense that
\be
{\partial\psi\over\partial \bar z} = [\psi,H],\quad
{\partial\psi^*\over\partial \bar z} = [\psi^*,H].
\en
We can then search for normal-ordered polynomials in $\psi,\psi^*$ and
their derivatives whose integrals commute with $H$; the first few conserved
currents are as follows:
\be
\psi^*\psi, \quad\psi^*\psi',\quad\psi^*\psi'' +
\kappa\psi^{*2}\psi^2,\quad\psi^*\psi'''+ 6\kappa\psi^*\psi''+
3\kappa\psi^{*2}\psi\psi'.
\en

We showed earlier that the \nls equation and the KP hierarchy were
related at the classical level, and we now extend that analysis to the
quantum case.  Just as the Poisson algebra \Wi\ has a realization in terms
of
classical scalar fields $\psi$ and $\psi^*$, the quantum algebra \Wi\ has a
realization in terms of operators $\psi$ and $\psi^*$ with OPE given by eqn
(\ref{ope}).  The quasiprimary fields $V^i$ of \Wi\ have the form
\be
V^i = {2n+2\choose n+1}^{-1}
\sum_{r=0}^{i+1} (-1)^r {i+1\choose r}^2 \psi^{(r)}\psi^{*(i+1-r)}.
\en
On substituting these expressions into the quantum KP charges $Q_i$ of the
previous section we recover the charges of the quantum \nls equation given
above, up to the freedom to add lower spin charges.  Some care must be
taken in transforming from the
normal ordering in terms of the $V^i$ to the free field normal ordering
used for the \nls charges.  Since the KP charges commute it follows that
the \nls charges will also commute, and because the \nls and KP charges are
uniquely determined by requiring that they commute with $H_3$ we conclude
that we obtain all the \nls charges in this way.

Acknowledgement: We are grateful to Klaus Hornfeck for discussions,
and to John Schwarz who specifically encouraged us to consider the
question of which theories possessed exceptional \W-algebras as symmetries.
M.D. Freeman is grateful to the UK Science
and Engineering Research Council for financial support.
%
%
%


\begin{thebibliography}{99}
%
%
\bibitem{DS}V.G. Drinfeld and V.V. Sokolov, J. Sov. Math. {\bf{30}} (1984)
1975.
\bibitem{Wi}G. Wilson, Ergod. Th. and Dynam. Sys. {\bf{1}} (1981) 361.
\bibitem{G}J.-L. Gervais, \PL B160 (1985) 277; \PL B160 (1985) 279.
\bibitem{FL}V.A. Fateev and S.L Lukyanov, Int. J. Mod. Phys. A3 (1988) 507;
``Additional symmetries and exactly-soluble models in two-dimensional
conformal field theory,'' Moscow preprint 1988.
\bibitem{SY}R. Sasaki and I. Yamanaka, Adv. Studies in Pure
Math. {\bf{16}} (1988) 271.
\bibitem{Z}A.B. Zamolodchikov, ``Integrable field theory from
conformal field theory,'' Proceedings of the Taniguchi Symposium, Kyoto,
(1988); Int. J. Mod. Phys. A3 (1988) 743; Int. J. Mod. Phys. A4 (1989)
4235.
\bibitem{EY1}T. Eguchi and S-K. Yang, \PL B224 (1989) 373.
\bibitem{HM}T. Hollowood and P. Mansfield, \PL B226
(1989) 73.
\bibitem{FF}B. Feigin and E. Frenkel, \PL B276 (1992) 79.
\bibitem{FW}M. D. Freeman and P. West,. \PL B295 (1992) 59.
\bibitem{Wa}Y. Watanabe, Lett. Math. Phys. 7 (1983) 99.
\bibitem{Ya}K. Yamagishi, \PL B259 (1991) 436.
\bibitem{YW}F. Yu and Y-S. Wu, \PL B263 (1991) 220.
\bibitem{PRS}C.N. Pope, L.J. Romans and X. Shen, \PL B242 (1990) 401.
\bibitem{W}S. Wolfram, ``Mathematica: a system for doing mathematics by
computer,'' Addison-Wesley, 1991.
\bibitem{FHW}M.D. Freeman, K. Hornfeck and P. West, ``Commuting quantities
and exceptional W-algebras,'' to be published in Int. J. Mod. Phys.
\bibitem{BBSS}F. Bais, P. Bouwknegt, M Surridge and K. Schoutens, Nucl. Phys.
B304 (1988) 348.
\bibitem{KW}H.G. Kausch and G.M.T. Watts, Nucl. Phys. B354 (1991) 740.
\bibitem{B}R. Blumehagen et al, Nucl. Phys. B361 (1991) 255.
\bibitem{Z2}A.B. Zamolodchikov, Teo. Mat. Fiz {\bf{65}} (1985) 347.
\bibitem{BPZ}A.A. Belavin, A.M. Polyakov, A.B. Zamolodchikov, Nucl. Phys. B241
(1984) 333.
\bibitem{K}H.G. Kausch, \PL B259 (1991) 448.
\bibitem{T}K. Thielemans, Int. J. Mod. Phys. {\bf{C2}} (1991) 787.
\bibitem{PRS1}C.N. Pope, L.J. Romans and X. Shen,\PL B236 (1990) 173;
\bibitem{PRS2}C.N. Pope, L.J. Romans and X. Shen, Nucl. Phys. B339 (1990) 191.
\bibitem{P}C.N. Pope, ``Lectures on W algebras and W gravity,'' preprint CTP
TAMU-103/91.
\bibitem{EY2}T. Eguchi and S-K. Yang, \PL B235 (1990) 282.
\bibitem{OSSY}M. Omote, M. Sakagami, R. Sasaki and I. Yamanaka, Phys. Rev.
D35 (1987) 2423.

%
\end{thebibliography}
\end{document}